\documentstyle[12pt,epic,eepic]{article}
\newcommand{\abs}[1]{\left\vert{#1}\right\vert}
\newcommand{\delm}{\partial_\mu}

\newcommand{\tr}{\hbox{tr}}
\newcommand{\bra}[1]{\left\langle{#1}\right|}
\newcommand{\ket}[1]{\left|{#1}\right\rangle}
\newcommand{\nonum}{\nonumber\\}
\newlength{\mleng}
\newcommand{\mq}{\makebox[0.5\mleng]{ }}
\newcommand{\mqq}{\makebox[\mleng]{ }}
\newcommand{\mqqq}{\makebox[1.5\mleng]{ }}

\renewcommand{\Im}{\hbox{Im}\;}
\renewcommand{\Re}{\hbox{Re}\;}
\newcommand{\be}{\begin{equation}}
\newcommand{\ee}{\end{equation}}
\newcommand{\ba}{\begin{eqnarray}}
\newcommand{\ea}{\end{eqnarray}}
\newcommand{\bd}{\begin{displaymath}}
\newcommand{\ed}{\end{displaymath}}

\newcommand{\ssize}{\scriptsize}

\newcommand{\Trang}[1]{\,\tr\left\langle{#1}\right\rangle}
\begin{document}

\begin{titlepage}

\begin{flushright}
\begin{minipage}[t]{3.5cm}
\begin{flushleft}
KUNS-1210 \\
HE(TH) 93/07\\
hep-ph/9307366 \\
July 27, 1993
\end{flushleft}
\end{minipage}
\end{flushright}

\vspace{0.7cm}

\begin{center}
\Large\bf
Axial Anomaly and Transition Form Factors
\end{center}

\vfill

\begin{center}
{\large
Masako {\sc Bando}}\footnote{
Permanent address: Aichi University, Miyoshi Aichi 470-02, Japan} 
\ and \ 
{\large Masayasu {\sc Harada}}%
\footnote{Fellow of the
Japan Society for the Promotion of Science for Japanese Junior
Scientists} \\
\ \\
{\normalsize\it
Department of Physics,
Kyoto University,\\
Kyoto 606-01, Japan}
\end{center}

\vfill

\begin{abstract}
We investigate the properties of the amplitude induced by the anomaly.
In a relatively high energy region 
those amplitudes are constructed by the vector meson poles
and the anomaly terms,
in which the anomaly terms can be essentially 
evaluated by the triangle quark graph.
We pay our attention to the anomaly term 
and make intensive analysis of the existing experimental data, i.e., 
the electromagnetic $\pi^0$ and $\omega$ transition form factors.
Our result shows that it is essential to use the constituent quark
mass instead of the current quark mass
in evaluating the anomaly term from the triangle graph.
\end{abstract}

\end{titlepage}

\begin{figure}[htbp]
\begin{center}
\setlength{\unitlength}{0.0050in}
\begin{picture}(385,419)(0,-10)
\thicklines
\put(180.000,275.000){\arc{40.000}{4.7124}{7.8540}}
\put(180.000,315.000){\arc{40.000}{1.5708}{4.7124}}
\put(180.000,355.000){\arc{40.000}{4.7124}{7.8540}}
\put(305.000,70.000){\arc{42.426}{3.9270}{7.0686}}
\put(335.000,40.000){\arc{42.426}{0.7854}{3.9270}}
\put(85.000,100.000){\arc{42.426}{2.3562}{5.4978}}
\put(55.000,70.000){\arc{42.426}{5.4978}{8.6394}}
\put(25.000,40.000){\arc{42.426}{2.3562}{5.4978}}
\put(275.000,100.000){\arc{42.426}{0.7854}{3.9270}}
\put(180,255){\blacken\ellipse{10}{10}}
\put(180,255){\ellipse{10}{10}}
\put(260,115){\blacken\ellipse{10}{10}}
\put(260,115){\ellipse{10}{10}}
\put(100,115){\blacken\ellipse{10}{10}}
\put(100,115){\ellipse{10}{10}}
\path(180,255)(100,115)(260,115)(180,255)
\path(210,325)(210,305)
\path(220,200)(220,185)
\path(220,185)(205,190)
\path(185,125)(170,115)(185,105)
\path(130,190)(145,195)(145,180)
\path(285,65)(300,50)
\path(300,50)(290,50)
\path(300,60)(300,50)
\path(90,50)(75,35)
\path(75,35)(85,35)
\path(75,45)(75,35)
\path(180,255)(100,115)(260,115)(180,255)
\path(180,255)(100,115)(260,115)(180,255)
\path(205,310)(210,305)
\path(215,310)(215,310)(210,305)
\put(175,390){\makebox(0,0)[lb]{\ssize$j_5^\mu$}}
\put(220,305){\makebox(0,0)[lb]{\ssize$k$}}
\put(265,35){\makebox(0,0)[lb]{\ssize$q$}}
\put(345,0){\makebox(0,0)[lb]{\ssize$j^\rho$}}
\put(0,0){\makebox(0,0)[lb]{\ssize$j^\nu$}}
\put(90,20){\makebox(0,0)[lb]{\ssize$p$}}
\put(265,35){\makebox(0,0)[lb]{\ssize$q$}}
\put(205,245){\makebox(0,0)[lb]{\ssize$\gamma^\mu\gamma_5$}}
\put(265,130){\makebox(0,0)[lb]{\ssize$\gamma^\rho$}}
\put(55,130){\makebox(0,0)[lb]{\ssize$\gamma^\nu$}}
\end{picture}
\ \\
\end{center}
\caption{The quark triangle anomaly graph
\label{fig:trianglegraph}}
\end{figure}
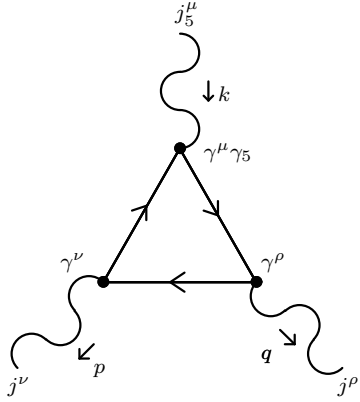

Much interests have been paid to the non-abelian anomaly 
because of its fundamental property of quantum gauge theories. 
These kinds of anomaly  are known to be caused by the  
quantum effects of triangle fermion loops
shown in Fig~\ref{fig:trianglegraph}. 
Their amplitudes in the low energy limit is known to be 
independent either of 
the higher order corrections  or of the internal fermion masses. 
On the other hand the asymptotic freedom of QCD facilitates us to 
predict the hadronic amplitudes at extremely high energy region, where 
again the triangle diagram becomes dominant. 
Indeed this triangle amplitude correctly reproduces
the high energy behavior obtained from the operator product expansion
(OPE) technique\cite{LepageBrodsky,Manohar:ZWpi:90},
up to the ambiguity of the coefficient factor.

The important observation here is that  
the amplitudes induced by the anomaly 
are constrained not only by the low energy theorem%
\cite{Adler:Anom:69,BellJackiw:Anom:69,AdlerBardeen}
but also by the  high energy behavior because 
of the asymptotically freedom,
and therefore we may expect that
the amplitudes induced by the anomaly are essentially controlled 
by the triangle quark graph.
The most well-known example of the processes induced by 
the axial anomaly is $\pi^0\rightarrow\gamma\gamma$.
The triangle quark amplitude
reproduces the experimental data
for this process excellently.

If we want to apply the anomaly induced amplitudes
to the relatively higher energy region,
the non-anomalous contributions come in the process in addition to 
the anomaly terms.
Then the QCD corrections to this triangle graph 
becomes important.
Especially 
we have to take account of the following 
nonperturbative QCD effects:
1) chiral symmetry breaking
(generation of constituent quark mass);
2) confinement effects;
3) hadron contributions.
Our interest here is investigating the processes 
induced by the axial anomaly in the high energy region.

In the separate paper\cite{BandoHarada:Anom1},
taking account of the above effects,
we have proposed an interpolating formula
for the amplitude induced by the axial anomaly.
There the amplitudes 
are constructed by the ``vector meson pole terms''
and the ``anomaly terms'',
the latter being essentially evaluated by the quark triangle graph
of Fig.~\ref{fig:trianglegraph}.

In this paper we pay our attention to the anomaly term 
and make intensive comparison with the existing experimental data.
The candidates for investigating the structure of 
this triangle anomaly amplitudes
are the $\pi^0$ transition form factors.

The result shows that,
in addition to the vector meson poles,
it is essential to use the constituent quark mass
instead of the current quark mass
for the internal quark lines of the triangle graph.

Let us start with the following three-point function:
\be
  T^{AB,\mu\nu\rho}(p,q)
  =
  - i \int d^4x d^4y e^{-i k\cdot y + i p\cdot x}
  \bra{0}\mbox{T} j^{A\nu}(x) j^{B\rho}(0) j_5^\mu(y) \ket{0},
\label{def:threepoint}
\ee
where $q\equiv k-p$,
$j_5^\mu$ is the axial-vector current
which generally couples to Nambu-Goldstone (NG) boson $P$:
$j_5^\mu = \bar{\psi} T^{\rm P} \gamma^\mu \gamma_5 \psi$,
$\bra{0}j_5^\mu(0)\ket{P(\mbox{\boldmath k})}= 
i k_\mu f_{\rm P}$,
and $j^{A\nu}(x)$ and $j^{B\rho}(0)$ are the relevant currents,
$j^{A\nu} = \bar{\psi} Q^A \gamma^\nu \psi$ 
with $Q^A$ being the charge matrix of the quark field $\psi$.
Denote the amplitude the NG-boson pole removed as 
$\widehat{T}^{AB,\mu\nu\rho}(p,q)$:
\be
  T^{AB,\mu\nu\rho}(p,q)
  =
  \widehat{T}^{AB,\mu\nu\rho}(p,q)
  + \frac{f_{\rm P} k^\mu}{m_{\rm P}^2-k^2} 
  {\cal T}^{AB,\nu\rho}(p,q),
\ee
where
\be
  {\cal T}^{AB,\nu\rho}(p,q) 
  = 
  - i \int d^4x e^{-ipx}
  \bra{0}\mbox{T} j^{A\nu}(x) j^{B\rho}(0) 
  \ket{P(\mbox{\boldmath k})} .
\ee
If one fixes the current $j^{A\nu}=j_{\rm e.m.}^\nu$,
the general form of this ${\cal T}^{AB,\nu\rho}(p,q)$ is given by%
\cite{GuberinaKuhnPecceiRuckel:80,%
ArnellosMarcianoParsa:ZWdecay:82}.
\ba
  {\cal T}^{AB,\nu\rho}(p,q)
&=&
  - \frac{N_{\rm c} e^2}{4\pi^2 f_{\rm P}}
  \Trang{ T^{\rm P} \left\{ Q^A , Q^B \right\} }
\nonum
&{}&
  \times
  \Biggl[
    ((p\cdot q) g^{\nu\rho} - q^\nu p^\rho) F_{\rm P}^{(5)}(p^2,q^2)
    {}+
    (p^2 g^{\nu\rho} - p^\nu p^\rho) G_{\rm P}^{(5)}(p^2,q^2)
\nonum
&{}& \mq
    {}+
    (p^2 q^\nu q^\rho - (p\cdot q) q^\nu p^\rho) H_{\rm P}^{(5)}(p^2,q^2)
    +
    \varepsilon^{\alpha\beta\nu\rho}
    p_\alpha q_\beta F_{{\rm P}}(p^2,q^2) 
  \Biggr] ,
\label{general:form}
\ea
where $N_{\rm c}$($=3$) is the number of colors and
we use the electromagnetic current conservation:
$p_\nu T^{AB,\nu\rho}(p,q) = 0$.
We have normalized in such a way that 
$F_{\rm P}(p^2=0,q^2=0)=1$ (see Eq.(\ref{def:piamp})).
If we further set $p^2=0$ (real photon),
we have the electromagnetic transition amplitude
$\bra{P(\mbox{\boldmath k})} j^{B\rho}(0) \ket{\gamma(p)}$,
which is expressed  in terms of 
the vector and axial form factors 
$F_{{\rm P}}(p^2,q^2)$ and $F^{(5)}_{{\rm P}}(p^2,q^2)$.
In the case where the current $j^{B\rho}(0)$ couples
to the photon or $Z$ boson,
the axial vector form factor vanishes,
and only the term proportional to 
$\varepsilon^{\alpha\beta\nu\rho}$ remains.
Thus the only anomaly term $F_{\rm P}(p^2,q^2)$
contributes to such processes.
In this paper, we consider this form factor $F_{{\rm P}}(p^2,q^2)$,
which is directly related to the anomaly part of this three-point
amplitude.

To see how this form factor is incorporated to the anomaly term,
we study the anomalous Ward-Takahashi identity
expressed as%
\cite{Hikasa:Anom:90,DeshpandePalOlness:90}
\be
  \delm j_5^\mu = 2 m_0 j_5 + \frac{e^2}{16\pi^2} F \widetilde{F},
\label{eq:WTidentity}
\ee
where $j_5$ is the corresponding pseudoscalar density:
$2m_0 j_5 \equiv 2 \bar{\psi} {\cal M} T^{\rm P} i \gamma_5 \psi$
(${\cal M}$: the mass matrix of the quarks).
This leads us to 
\be
  k_\mu T^{AB,\mu\nu\rho}(p,q)    
  =
  M^{AB,\mu\nu} (p,q) 
  +  \frac{N_{\rm c} e^2}{4\pi^2} 
  \Trang{ T \left\{ Q^A , Q^B \right\} }
  \varepsilon^{\alpha\beta\nu\rho} p_\alpha q_\beta,
\label{eq:anomalousWT}
\ee
where
\be
  M^{AB,\nu\rho}(p,q)
  =
  \int d^4x d^4y e^{-ipx+ik}
  \bra{0}\mbox{T} j^{A\nu}(x) j^{B\rho}(0) 2 m_0 j_5 (y) \ket{0}.
\ee
This amplitude $M^{AB,\nu\rho}(p,q)$,  
as well as 
the three-point function $T^{AB,\mu\nu\rho}(p,q)$, 
contains the NG-boson pole contributions.
We extract the NG-boson pole contributions in both sides:
\ba
&{}&
  \left[
    k_\mu \widehat{T}^{AB,\mu\nu\rho}(p,q)
    + \frac{f_{\rm P} k^2}{m_{\rm P}^2-k^2} 
    {\cal T}^{AB,\nu\rho}(p,q)
  \right]
\nonum
&=&
  \left[
    \widehat{M}^{AB,\nu\rho}(p,q) +
    \frac{f_{\rm P} m_{\rm P}^2}{m_{\rm P}^2-k^2} 
    {\cal T}^{AB,\nu\rho}(p,q) 
  \right]
  +  \frac{N_{\rm c} e^2}{4\pi^2} 
  \Trang{ T \left\{ Q^A , Q^B \right\} }
  \varepsilon^{\alpha\beta\nu\rho} p_\alpha q_\beta,
\label{eq:anomalousWT2}
\ea
where use has been made of 
$\bra{0}2m_0j_5(0)\ket{{\rm P}}=m_{\rm P}^2 f_{\rm P}$.
Thus the transition amplitude for NG-boson
is obtained:
\be
  {\cal T}^{AB,\nu\rho}(p,q) 
  =
  \frac{1}{f_{\rm P} }
  \left[
    k_\mu \widehat{T}^{AB,\mu\nu\rho}(p,q)
    - \widehat{M}^{AB,\mu\nu} (p,q) 
  \right]
  -  \frac{N_{\rm c} e^2}{4\pi^2 f_{\rm P}} 
  \Trang{ T \left\{ Q^A , Q^B \right\} }
  \varepsilon^{\alpha\beta\nu\rho} p_\alpha q_\beta.
\label{eq:anomalousWardTakahashi}
\ee
which is rewritten in terms of the invariant amplitude
\ba
  {\cal T}^{AB,\nu\rho}(p,q) 
&=&
  \frac{N_{\rm c}e^2}{4\pi^2f_{\rm P}} 
  \Trang{T \left\{ Q^A ,  Q^B \right\} }
  \varepsilon^{\alpha\beta\nu\rho} p_\alpha q_\beta
  \Biggl[
    1 - T(p^2,q^2,k^2)
  \Biggr] 
\nonum
&{}&
  -  \frac{N_{\rm c} e^2}{4\pi^2 f_{\rm P}} 
  \Trang{ T \left\{ Q^A , Q^B \right\} }
  \varepsilon^{\alpha\beta\nu\rho} p_\alpha q_\beta.
\label{def:piamp}
\ea
{}From this expression 
we easily see that $T(p^2,q^2,k^2)$ corresponds to
$F_{\rm P}(p^2,q^2)$ of Eq.(\ref{general:form}).
It is well known 
that the triangle diagram containing fermion loops
couples to vector or axial vector currents leads to anomalies,
and in the standard renormalization procedure
the gauge invariance guarantees that 
any  higher order corrections do not
modify the structure of this anomaly.

So let us first calculate the quark triangle graph shown in
Fig.~\ref{fig:trianglegraph} 
which is expected to give the dominant contribution
to the function $T(p^2,q^2,k^2)$ in Eq.(\ref{def:piamp}).
The result is expressed as%
\footnote{Here we restrict ourselves to neutral currents.}
\be
  \left.
    T(p^2,q^2,k^2)
  \right\vert_{\rm triangle}
  =
  \frac{
    2 \sum_i T_i Q^A_i Q^B_i
  }{
    \Trang{T \left\{ Q^A,  Q^B \right\} }
  }  
  \int [dz] 
  \frac{m_i^2}{m_i^2 - z_2z_3k^2 - z_3z_1p^2 - z_1z_2 q^2},
\label{res:functionT}
\ee
with the Feynman parameter integral defined by
$
  \int [dz]
  \equiv
  2 \int_0^1
  dz_1 dz_2 dz_3 \delta (1-z_1-z_2-z_3) 
$.
It is easy to see that in the low energy limit in which
$k^2=p^2=q^2=0$, this reduces to
\be
  \left.
    T(p^2=0,q^2=0,k^2=0)
  \right\vert_{\rm triangle}
  =
  1,
\label{eq:normalization}
\ee
independently of $m_i$.
This is the expression that the anomaly term is exactly reproduced by the
lowest order triangle graph.

At this point,
we should stress that 
the quark mass $m_i$ appearing in $T(p^2,q^2,k^2)$ is
taken to be the {\it constituent quark masses}
instead of the original current quark masses.
This reflects the most prominent feature of QCD:
since the chiral symmetry has been spontaneously broken 
and there appear NG-bosons,
the quarks necessarily acquire dynamical masses.
This replacement never changes the value of the low energy limit, 
because $T(p^2=0,q^2=0,k^2=0)=1$ independently of the quark mass 
(see Eq.(\ref{eq:normalization})).

In the following we restrict ourselves to the $\pi^0$ case,
where only the $u$ and $d$ quarks come into play.
Then $j_5^\mu = \sum_i \bar{q}_i T_i \gamma^\mu \gamma_5 q_i$,
$q_i=(u,d)$, $T_i=(1/2,-1/2)$,
and $2m_0 j_5 \equiv \sum_i 2 m_{0i} \bar{q}_i T_i i \gamma_5 q_i$.
The triangle graph contribution to $T(p^2,q^2,k^2=0)$ is obtained as
\be
  I(p^2,q^2;m) 
  \equiv
  \left.
    T(p^2,q^2,k^2=0)
  \right\vert_{\rm triangle} 
  =
  \int [dz] 
  \frac{m^2}{m^2 - z_3z_1p^2 - z_1z_2 q^2} ,
\label{def:functionI}
\ee
where $m$ (=$m_u\simeq m_d$) is the constituent quark mass
(of the $u$ and $d$ quraks).

If one photon is on its mass shell (e.g., $p^2=0$), then the function
$I(p^2,q^2;m)$ reduces to 
\ba
  J(q^2;m)
&\equiv&
  I(p^2=0,q^2;m) 
\nonum
&=&
  - \frac{m^2}{q^2}
  \left[
    \left\{
      \ln
      \frac{
        \sqrt{4m^2-q^2} + \sqrt{-q^2}
      }{
        \sqrt{4m^2-q^2} - \sqrt{-q^2}
      }
    \right\}^2
  \right] ,
\label{def:tildeI}
\ea
which gives the $\pi^0\gamma$ transition form factor.
The high energy behavior of this function is given by
\be
  J(q^2; m)
  \mathop{\Rightarrow}_{q^2\gg m^2}
  \frac{m^2}{q^2}
  \left[
    \pi^2 - 
    \left(
      \ln \frac{m^2}{q^2}
    \right)^2
    - 2 i \pi \ln \frac{m^2}{q^2}
  \right] .
\label{eq:Itilde}
\ee

It is well known that, in the low energy limit,
the higher order corrections do not change 
the value of the three-point function induced by the anomaly 
in Eq.(\ref{def:threepoint}), or $T(p^2=q^2=k^2=0)$ itself.
However, QCD corrections do 
contribute appreciably to its higher energy
behavior.
Especially nonperturbative QCD effects are alleged to take place:
1) chiral symmetry breaking
(generation of constituent quark mass);
2) confinement effects;
3) hadron contributions.

As we have mentioned below Eq.(\ref{eq:normalization}),
we have already taken account of the effect 1) 
by replacing the current quark masses by the constituent masses of the
quarks.
The next important QCD effects are
represented as a rich hadron spectrum.
As an average, most part of those resonances with broad widths 
may be already involved by the above replacement
in the triangle contributions of Fig.~\ref{fig:trianglegraph}.

\begin{figure}[htbp]
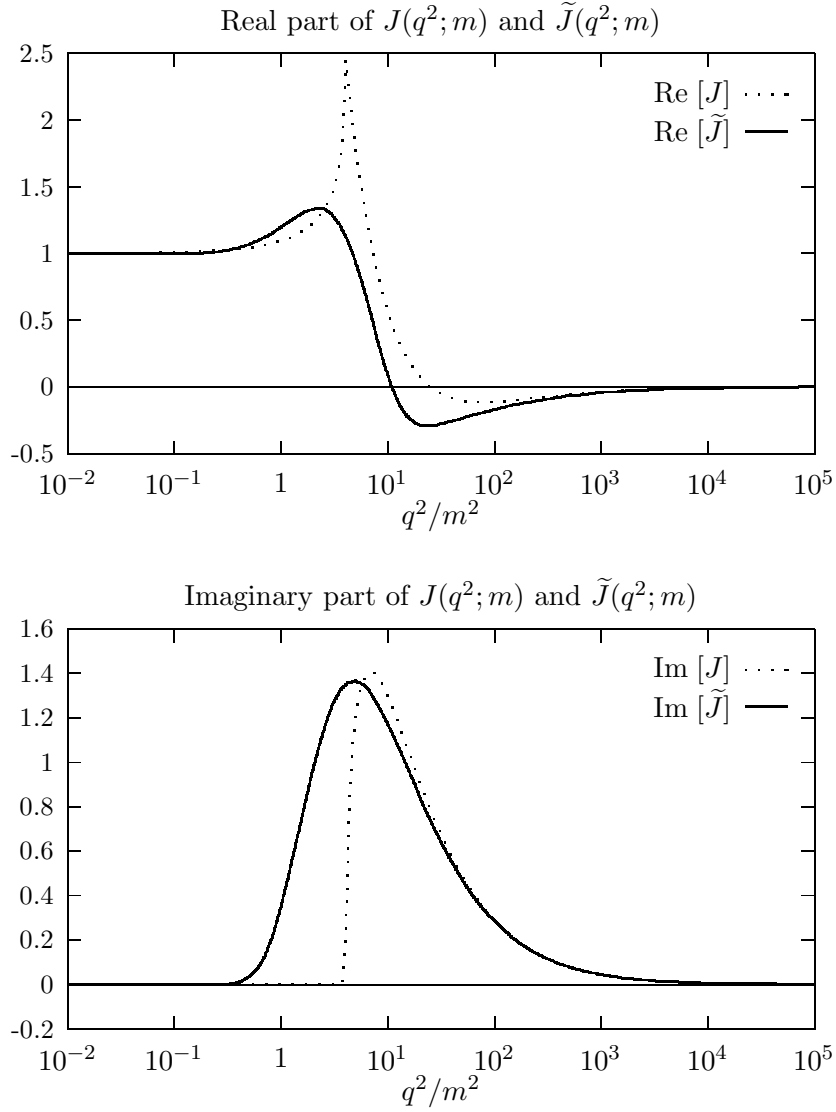

\begin{center}
\input{modfunre}
\input{modfunim}
\end{center}
\caption[]{The QCD corrected-improved 
functions $\widetilde{J}(q^2;m)$ (solid lines)
comparing with the original functions $J(q^2;m)$ (dotted lines), 
where we take the parameter choice 
$\alpha=1.4$, $\beta=0.35$ and $\lambda=2.5$.
\label{fig:modifiedI}}
\end{figure}

However, one should further modify the functions
$I(p^2,q^2;m)$ and $J(q^2;m)$
due to the effect 2).
The graph in Fig.~\ref{fig:trianglegraph} shows the contribution of
the $q\bar{q}$ threshold at $p^2=4m^2$ ($q^2=4m^2$) 
from which an imaginary part emerges
as shown in Fig.~\ref{fig:modifiedI} (the dotted lines).
However, since the quarks are confined,
the intermediate states are not multi-quark states 
but actually multi-hadron states
(2$\pi$, 4$\pi$, ... etc.).
The above effect may be properly taken into account by smearing out
the original $J(q^2;m)$.
Here we adopt the following function:
\be
  \widetilde{J}(q^2;m)
  \equiv
  - \frac{\left(\ln(1+x)\right)^2}{x}
  + \frac{\ln(1+x)}{x} + \alpha x \exp(-\beta x)
  + 2 \pi i \frac{\ln(1+x)}{x} 
  \exp\left(-\frac{\lambda}{x} \right) ,
  \, 
  x=\frac{q^2}{m^2}  .
\label{smooth:tildeI}
\ee
Both $J$ and $\widetilde{J}$ are 
shown in Fig.~\ref{fig:modifiedI},
where we take the parameter choice 
$\alpha=1.4$, $\beta=0.35$ and $\lambda=2.5$.
{}From this figure, we see that the modified amplitudes
$\widetilde{J}$ are properly smeared out.
This form has been chosen so as to 
coincide with the original $J$ 
(the dotted lines)
both in the high and low energy limits.

Finally, we must take account of the dominant hadron poles 
as the effect 3).
Among hadron spectrum, NG-bosons,
which have been put in as elementary fields,
are of course most important.
The next important poles are lowest energy bound states,
i.e., the vector mesons.
This can be done by introducing the vector meson poles to the
amplitude,
where we should be careful to include them 
to avoid the double counting
and to satisfy
the low energy theorem Eq.(\ref{eq:normalization}).

Taking all the above mentioned corrections into account,
we propose a probable expression of form factors.
The $\pi^0 \gamma^{\ast} \gamma^{\ast}$ transition form factor is%
\footnote{%
The function $\widetilde{I}$ is obtained by the same procedure as
$\widetilde{J}$.}
\ba
  F_{\pi^0\gamma^{\ast}\gamma^{\ast}} (p^2,q^2)
&=&
  \kappa_1 \widetilde{I}(p^2,q^2;m) 
\nonum
&{}&
  {}+ \kappa_2
  \left[
    \left\{
      D_\rho(p^2) + D_\omega(p^2)
    \right\}
    \widetilde{J}(q^2;m) +
    \left\{
      D_\rho(q^2) + D_\omega(q^2)
    \right\}
    \widetilde{J}(p^2;m)
  \right]
\nonum
&{}& 
  {}+ \kappa_3
  \left[
    D_\rho(p^2) D_\omega(q^2)
    + D_\rho(q^2) D_\omega(p^2)
  \right] ,
\label{def:formfactor}
\ea
where $D_\rho(p^2)$ and $D_\omega(p^2)$ are 
the Breit-Wigner type propagators for $\rho$ and $\omega$ mesons:
\be
  D_\rho(p^2)
  =
  \frac{m_\rho^2}{m_\rho^2-p^2 - i \sqrt{p^2}\Gamma_\rho} ,
\mqq
  D_\omega(p^2)
  =
  \frac{m_\omega^2}{m_\omega^2-p^2 - i \sqrt{p^2}\Gamma_\omega} ,
\ee
with $\Gamma_\rho$ and $\Gamma_\omega$ being
the widths of $\rho$ and $\omega$ mesons, respectively.
The weight of $D_\rho$ and $D_\omega$ has been 
chosen to be consistent with the flavor symmetry.
Since the above form factor should satisfy the low energy theorem,
the normalization condition Eq.(\ref{eq:normalization}) gives
$\kappa_1+4\kappa_2+\kappa_3=1$
(note that $D_\rho(p^2=0)=D_\omega(p^2=0)=1$).
This expression is most elegantly derived
in the framework of hidden local symmetry
(HLS) (see, for a review, Ref.\cite{BandoKugoYamawaki:PRep}).
In the following we use the parameters $c_3$ and $c_4$:
$c_3=2\kappa_3$; $c_4=1-\kappa_1$,
where $c_3$ and $c_4$ 
correspond to the terms appearing in the Lagrangian 
constructed in the framework of HLS.
Then of course, this form factor again 
satisfies the low energy theorem:
$
  F_{\pi^0\gamma^{\ast}\gamma^{\ast}}
  (p^2=0,q^2=0) =1 
$,
independently of the parameters $c_3$ and $c_4$.
This together with the coefficient factor 
$g_{\gamma\gamma\pi}=-e^2/(4\pi^2f_\pi)$,
excellently reproduces the experimental value of the
$\pi^0\rightarrow\gamma\gamma$ process:
this is nothing but the expression of the low energy theorem.
In the high energy region, on the other hand,
in addition to the $\widetilde{J}(q^2;m)\sim (m^2/q^2)$
contribution,
we have terms
$D_\rho(q^2) \sim m_\rho^2/q^2$ and
$D_\omega(q^2) \sim m_\omega^2/q^2$.
If one notices that 
$m_\rho \sim m_\omega \sim {\cal O}(m)$,
it is easy to see that $\rho$ and $\omega$ pole
contributions have the same high energy properties as $\widetilde{I}$
and $\widetilde{J}$.

Now, we compare the above proposed form factor 
with the experimental data
and examine the $q^2$ dependence of the anomaly terms 
$\widetilde{I}$ and $\widetilde{J}$.
To do this, 
we first determine the parameter $c_3$ and $c_4$ from the experiment.

By setting $p^2=0$ we have the $\pi^0$ electromagnetic transition form
factor:
\be
  F_{\pi^0\gamma}(q^2)
  =
  \left(
    1-\frac{c_3+c_4}{2}
  \right)
  \widetilde{J}(q^2;m)
  + \frac{c_3+c_4}{4}
  \left\{
    D_\rho(q^2) + D_\omega(q^2)
  \right\} ,
\label{def:pigammaformfactor}
\ee
in which $(c_3+c_4)/2$ is 
to be determined.
The comparison of our formula with the experimental
data\cite{CELLO:91} indicates 
\be
  \frac{c_3+c_4}{2}
  =
  1.0 .
\label{fit:c3c4value}
\ee
This value implies that the vector meson dominance (VMD)
is almost realized in this process.
The result Eq.(\ref{fit:c3c4value}) implies that
the $q^2$ dependence of $\widetilde{J}$ is not so important for
the $\pi^0\gamma$ transition form factor, 
so far as we analyze the existing data
(see Ref.~\cite{BandoHarada:Anom1}).
So we can not get the information of the function
$\widetilde{J}(q^2;m)$ from this process alone.

On the other hand,
the vector meson form factors can also be obtained
from our formula.
The expression for the $\omega\pi^0$%
\footnote{
We could also obtain the $\rho^0\pi^0$ transition form factor.
However, the branching ratio
$B(\rho^0\rightarrow\pi^0\gamma)$ ($=(7.9\pm2.0)\times10^{-4}$) 
is far smaller than 
$B(\omega\rightarrow\pi^0\gamma)$ ($=(8.5\pm0.5)\times10^{-2}$).
Moreover, no experimental data for 
$\rho^0\rightarrow\pi^0\gamma^{\ast}$ has been reported.}
transition form factor is given by extracting 
the $\omega$-pole contributions from
Eq.(\ref{def:formfactor})
and by making proper normalization of the
amplitude we find
\be
  \mbox{(case 1)} \mqq
  F_\omega(q^2)
  =
  - \tilde{c} \widetilde{J}(q^2;m) + (1+\tilde{c}) D_\rho(q^2) ,
  \mqq
  \tilde{c} \equiv \frac{c_3-c_4}{c_3+c_4} \, .
\label{def:OmegaPiFormfactor}
\ee
where $\tilde{c}$ is to be determined.
We take here $m=m_\rho/2$ (case 1).
For reference, we examine the case 
where we adopt the current quark mass $m=m_0\simeq5$MeV
in $\widetilde{J}(q^2;m_0)$ (case 2),
and 
the case 
in which the triangle anomaly term is taken to be a constant
(case 3):
\ba
  \mbox{(case 2)} 
&\mqq&
  F_\omega^{(m_0)}(q^2)
  =
  - \tilde{c} \widetilde{J}(q^2;m_0) + (1+\tilde{c}) D_\rho(q^2) \, ,
\label{form:current}
\\
  \mbox{(case 3)} 
&\mqq&
  F_\omega^{\rm C}(q^2)
  =
  - \tilde{c} + (1+\tilde{c}) D_\rho(q^2) \, .
\label{form:BW0}
\ea
The form factor of case 3 was often used in the extensive analysis
of the form factors in relatively lower $q^2$ region 
(see, for example Ref.\cite{BramonGrauPancheri:ChPT:92}).

Let us determine the value of $\tilde{c}$ for each case
using the experimental data 
$\Gamma(\omega\rightarrow\pi^0\mu^+\mu^-)$.
It is convenient to use the following expression:
\ba
  \frac{
    \Gamma(\omega\rightarrow\pi^0\mu^+\mu^-)
  }{
    \Gamma(\omega\rightarrow\pi^0\gamma)
  }
&=&
  \int_{4 m_\mu^2}^{(m_\omega-m_\pi)^2} dq^2
  \frac{\alpha}{3\pi}
  \frac{1}{q^2}
  \left(
    1 + \frac{2m_\mu^2}{q^2}
  \right)
  \sqrt{\frac{q^2-4m_\mu^2}{q^2}}
\nonum
&{}& \mq 
  \times
  \left[
    \left(
      1 + \frac{q^2}{m_\omega^2-m_\pi^2}
    \right)^2
    - \frac{4m_\omega^2q^2}{(m_\omega^2-m_\pi^2)^2}
  \right]^{3/2}
  \abs{F_\omega(q^2)}^2 ,
\ea
where $q^2$ is the intermediate photon momentum 
(or invariant mass of final muons).
The experimental data\cite{ParticleDataGroup} shows
\be
  \frac{
    \Gamma(\omega\rightarrow\pi^0\mu^+\mu^-)
  }{
    \Gamma(\omega\rightarrow\pi^0\gamma)
  }
  =
  \frac{
    (9.6\pm2.3) \times 10^{-5} 
  }{
    (8.5\pm0.5) \times 10^{-2} 
  }
  \simeq
  (1.1\pm0.3) \times 10^{-3} ,
\ee
which gives%
\footnote{
The another solutions for $\tilde{c}$, for example, 
$\tilde{c} = -2.5 \pm 0.1$ (case 1), are
excluded by experiments.}
\ba
  \mbox{(case 1)} \mqqq
  \tilde{c}
&=&
  0.49 ^{\displaystyle +0.12}_{\displaystyle -0.13} ,
  \mqq
  \mbox{for} \mq F_\omega(q^2) ,
\nonum
  \mbox{(case 2)} \mqqq
  \tilde{c}
&=&
  0.11 ^{\displaystyle +0.16}_{\displaystyle -0.18} ,
  \mqq
  \mbox{for} \mq F_\omega^{(m_0)}(q^2) ,
\nonum
  \mbox{(case 3)} \mqqq
  \tilde{c}
&=&
  0.88 ^{\displaystyle +0.16}_{\displaystyle -0.18} ,
  \mqq
  \mbox{for} \mq F_\omega^{\rm C}(q^2) .
\label{value:ctilde}
\ea
The average values for all the above cases 
show that the complete $\rho$ meson dominance is 
incapable of describing
the $\omega\pi^0$ form factor\cite{Dzhelyadinetal1}
(VMD corresponds to $\tilde{c}=0$, see
Eq.(\ref{def:OmegaPiFormfactor})).

Now that
we have found that 
the $\omega\pi^0$ transition form factor 
receives appreciable contributions from the anomaly term,
we use the $\omega\pi^0\gamma^{\ast}$ process to get the information 
on the $q^2$ dependence of the anomaly term 
from the experimental data.
The available experimental data for $\omega\pi^0\gamma^{\ast}$
lies in a relatively wide energy range 
(up to $1.4$GeV).
We use the value of $\tilde{c}$ for each case obtained above,
and compare the above three cases.
The predicted curves are shown in Fig.~\ref{fig:formgl}.
In those figures
the three data points around $q^2=0.4$GeV${}^2$
are located far from the theoretical curves.
We should remark that 
the deviation of the above data points 
should not be taken seriously:
because the above region 
corresponds to the kinematical boundary,
the event rate is very rare.
As a result, 
the behavior of the form factor strongly depends on the 
normalization uncertainty\cite{Dzhelyadinetal1}.
{}From Fig.~\ref{fig:formgl} we find that
the form factor $F_\omega^{\rm C}(q^2)$ (case 3)
can not reproduce the high $q^2$ data,
the theoretical values are much
larger than the data points.
For the case 2 ($F_\omega^{(m_0)}(q^2)$)
the suppression of the form factor is too strong
and the theoretical values are
too small to fit the data points
in the high $q^2$ region.
On the other hand,
aside from the three points mentioned above,
the improved form factor $F_\omega(q^2)$ (case 1)
is found to be in good agreement with 
the data in the whole energy region.
This successful result is 
owing to the triangle anomaly term 
described by the modified function $\widetilde{J}(q^2;m)$
in addition to the vector meson pole effect.
In particular,
in the above analysis 
we have taken the constituent quark mass $m=m_\rho/2$.

\begin{figure}[htbp]
\begin{center}
\input{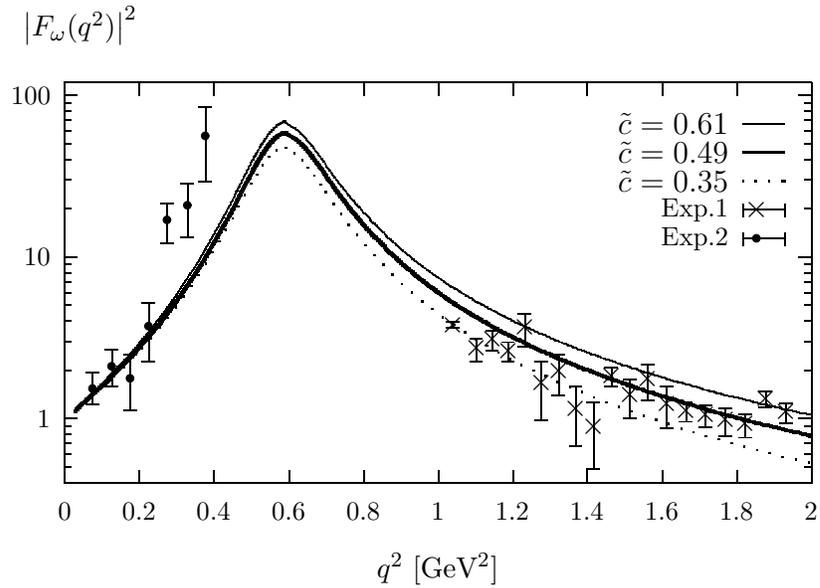}
\end{center}
\caption[]{
The form factor in the energy range below 1.4 GeV
using the unsmeared function $J(q^2;m)$.
\label{fig:formgli}
}
\end{figure}

One might wonder whether the above results
depend on the smearing procedure of the function $J$.
To check this we show the curve obtained from the form factor
with the unsmeared function $J$ in Fig~.\ref{fig:formgli}.
{}From this figure, we see that the smearing procedure does not affect
the value of the form factor so far as in the region 
where the experimental data exists,
although it does around the threshold ($q^2\simeq0.6$GeV${}^2$).

In conclusion, the experimental data 
in a relatively wide energy range is found to be reproduced 
very well by using the QCD corrected triangle amplitude
as the ``anomaly term''.
It must be stressed that the {\it constituent quark mass} 
plays an essential role in the triangle anomaly term.

It is a common understanding that 
the low-energy effective Lagrangian 
for quarks and gluons is described by the chiral Lagrangian in 
terms of the NG-bosons. 
Their anomaly term has been introduced 
by the Wess-Zumino action\cite{WessZumino,Witten:anomaly}
in the effective Lagrangian. 
Another way to incorporate the anomaly term
is to introduce the quarks 
(to be more exact, we should say ``constituent quarks")
and gluon fields in place of the hadrons other than NG-bosons 
in the chiral Lagrangian\cite{ManoharGeorgi}.

At extremely low energy limit,
the above two effective Lagrangians work very well 
for reproducing the phenomena 
and they are completely equivalent
because the latter induces the anomaly term via loop effects. 
This is the result of low energy theorem. 
If one wants to apply the Lagrangian 
to nonzero but relatively lower energy scale,
the vector mesons should be properly included in the Lagrangian.
So far as we examine the processes in which the complete VMD is
realized, 
the whole anomaly terms are replaced by the vector meson pole terms.
Indeed we have seen that the anomaly term disappears 
from the $\pi^0\gamma$ transition form factor 
and we cannot check the $q^2$ dependence of the anomaly term
from this process.
On the other hand, we have found that $\omega\pi^0$ transition form 
factor, in which the complete VMD hypothesis is not valid, recieves 
appreciable contributions from the anomaly term. 
For such processes the above two approaches
give different predictions.
Our result indicates that the WZ action alone 
cannot reproduce the experimental data for the $\omega\pi^0$
transition form factor in the relatively high enenrgy region,
because the WZ action gives the constant anomaly term.
This is in remarkable contrast to the anomaly term 
calculated from the QCD corrected quark triangle graph. 
In this sense our treatment here may have somewhat similar spirit as 
the latter approach.

Also the high energy behavior of the anomaly term,
with the usage of the constituent quarks in the internal loop,
seems to suggest the possibility that
this anomaly term can be applicable 
even to the phenomena in the extremely high energy region, 
say for examples, $Z\rightarrow P \gamma $
($q^2=m_Z^2$) or $W$ decay processes%
\cite{Manohar:ZWpi:90,GuberinaKuhnPecceiRuckel:80,%
ArnellosMarcianoParsa:ZWdecay:82,Hikasa:Anom:90,%
DeshpandePalOlness:90}.
The processes induced by the axial anomaly 
may yet provide us with the information on 
how to construct the effective theory applicable to hegher energy 
regions.

\vspace{0.3cm}

We would like to thank Taichiro Kugo 
for comments  and instructive discussions
about the property of the anomaly.

\begin{figure}[htbp]
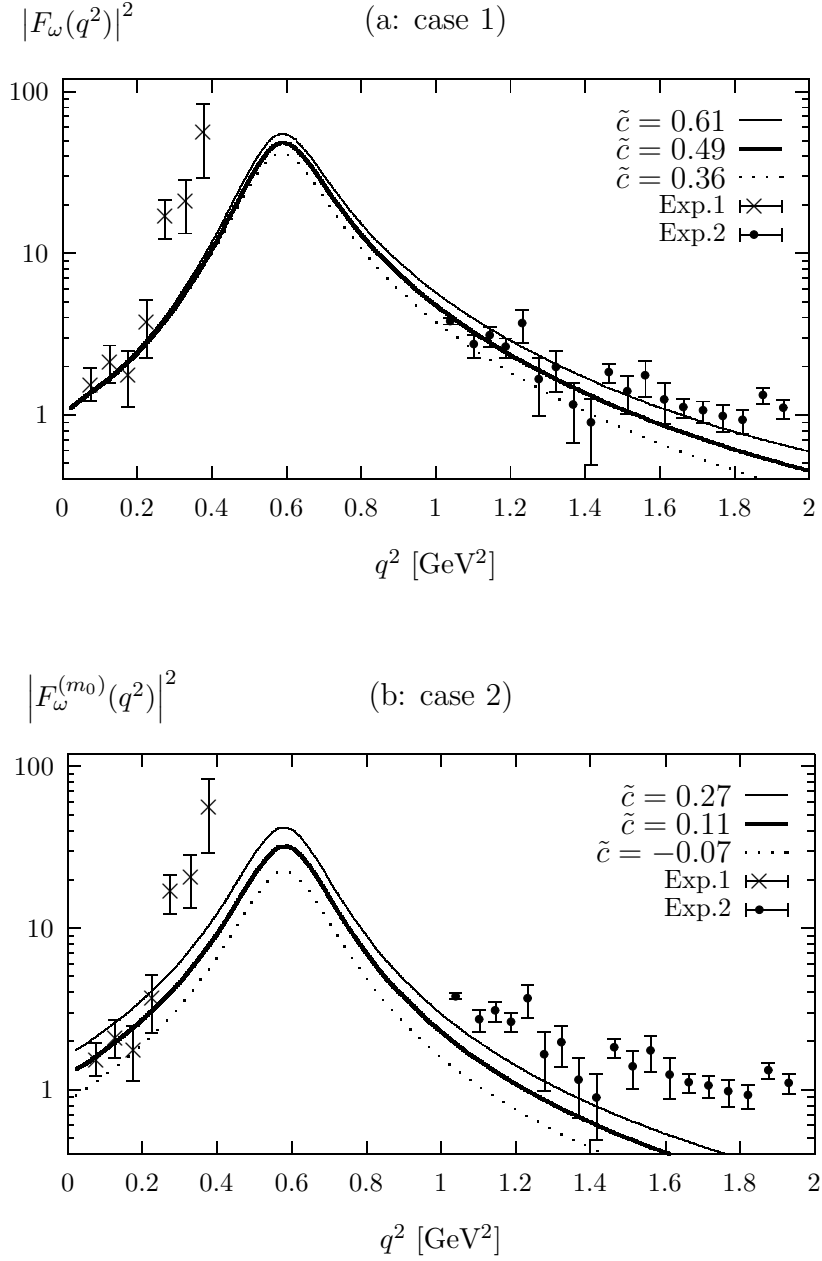

\begin{center}
\input{formgj}
\ \\
\ \\
\ \\
\input{formgj2}
\end{center}
\renewcommand{\thefigure}{\arabic{figure}~(a)~(b)}
\caption[]{
The form factors in the energy range below 1.4 GeV:
(a: case 1) the improved form factor $F_\omega(q^2)$;
(b: case 2) $F_\omega^{(m_0)}(q^2)$.
The low energy data (Exp.1) is given in Ref.\cite{Dzhelyadinetal1},
and the high energy data (Exp.2) is translated 
from the cross section data\cite{Dolinskyetal}.
The values of $\tilde{c}$ are determined using 
$\Gamma(\omega\rightarrow\pi^0\mu^+\mu^-)$
(see Eq.(\ref{value:ctilde})).
\label{fig:formgl}
}
\end{figure}

\newpage

\begin{figure}[htbp]
\begin{center}
\input{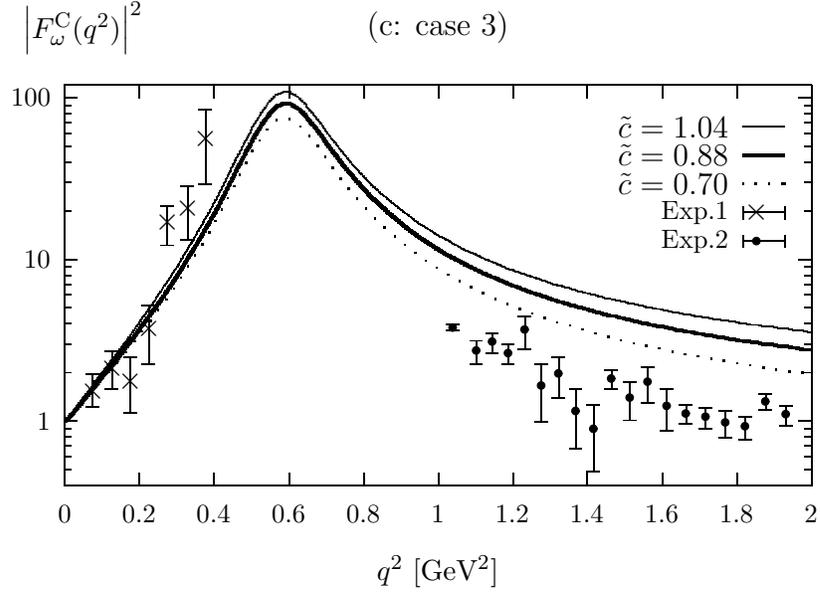}
\end{center}
\setcounter{figure}{2}
\renewcommand{\thefigure}{\arabic{figure}~(c)}
\caption[]{
The form factors in the energy range below 1.4 GeV:
(c: case 3) $F_\omega^{\rm C}(q^2)$.
}
\end{figure}

\end{document}